\documentclass[a4paper,11pt]{article}
\pdfoutput=1 

\usepackage{jinstpub} 

\usepackage[utf8]{inputenc}
\usepackage{natbib}
\usepackage{graphicx}
\usepackage{svg}
\usepackage{amsmath}
\usepackage{ulem} 
\usepackage{siunitx}

\usepackage{float}
\usepackage{parcolumns}
\usepackage{qtree}
\usepackage{url}
\newcommand{\ts}{\textsuperscript}

\usepackage{color}
\usepackage{xcolor}
\usepackage{caption}
\usepackage{textcomp}
\usepackage{todonotes}
\usepackage{listings}

\title{\boldmath Strategies for on-chip digital data compression for X-ray pixel detectors}

\author[a]{M. Hammer,}
\author[a]{K. Yoshii,}
\author[a]{A. Miceli\note{Corresponding author.}}

\affiliation[a]{Argonne National Laboratory,\\9700 S. Cass Ave., Lemont, IL 60439, U.S.A.}

\emailAdd{amiceli@anl.gov}

\abstract{
The continued desire for X-ray pixel detectors with higher frame rates will stress the ability of application-specific integrated circuit (ASIC) designers to provide sufficient off-chip bandwidth to reach continuous frame rates in the \SI{1}{\MHz} regime. To move from the current \SI{10}{\kHz} to the \SI{1}{\MHz} frame rate regime, ASIC designers will continue to pack as many power-hungry high-speed transceivers at the periphery of the ASIC as possible. In this paper, however, we present new strategies to make the most efficient use of the off-chip bandwidth by utilizing data compression schemes for X-ray photon-counting and charge-integrating pixel detectors. In particular, we describe a novel in-pixel compression scheme that converts from analog to digital
converter units to encoded photon counts near the photon Poisson noise level and achieves a compression ratio of $>\!1.5\times$ independent of the dataset. In addition, we describe a simple yet efficient zero-suppression compression scheme called ``zeromask'' (ZM) located at the ASIC's edge before streaming data off the ASIC chip. ZM achieves  average compression ratios of $>\!4\times$, $>\!7\times$, and $>\!8\times$ for high-energy X-ray diffraction, ptychography, and X-ray photon correlation spectroscopy datasets, respectively. We present the conceptual designs, register-transfer level block diagrams, and the physical ASIC implementation of these compression schemes in \SI{65}{\nm} CMOS. When combined, these two digital compression schemes could increase the effective off-chip bandwidth by a factor of 6--12$\times$. 

}


\arxivnumber{2006.02639}

\begin{document}
\maketitle
\flushbottom

\section{Introduction}
\label{sec:intro}

X-rays have unique potential for nanoscale resolution imaging of 
centimeter-sized objects \cite{ChrisBook}. Pushing X-ray imaging into the
nanoscale is crucial for understanding complex hierarchical systems on length
scales from the atomic scale to the macroscale, 
in order to address scientific questions ranging from materials 
science and biology to mechanical and civil engineering. The fourth-generation
storage ring light sources will increase X-ray beam brightness and coherent 
flux by 100 to 1,000 times over current values, with great advantages for 
science. Increases in nanoscale-focused brightness have motivated beamline 
development~\cite{Deng:2019es} of very fast scanning microscopy instruments 
with scanning rates approaching \SI{1}{\MHz} (i.e., \SI{1}{\us} dwell times) 
in order to be able to image large samples. In addition, high-energy X-ray 
diffraction \cite{Ren2012, Park:2017hb} provides nondestructive and in situ 3D
atomic and mesoscale information about structure and its evolution in the 
broad class of single-crystal and polycrystalline materials.

While the brightness increases provided by the accelerator and optics upgrades
will be essential, these beamlines 
and techniques will not reach their full potential if they are limited to 
using detectors of the type available 
today or those currently seen on the horizon. In particular, achieving full 
potential requires advancing the state of the art from 
present few-kilohertz frame rates to megahertz frame rates. Photon-counting 
detectors are fundamentally unable to provide high 
dynamic range per pixel at megahertz frame rates, especially given the bunched
time structures of storage rings and free electron lasers.  Full exploitation 
of bright X-ray light sources requires use of 
charge-integrating detectors to preserve high dynamic range at fast frame 
rates. A number of charge-integrating detector 
platforms \cite{MM-PAD,  JUNGFRAU, ePix10k} reach 
toward the required dynamic range but 
achieve frame rates of a few 
tens of kilohertz at best. At the other extreme, 
burst-mode detectors \cite{KeckPAD, AGIPD, LPD, DSSC, pixfel,SOPHIAS} are able to 
frame at  5--10 MHz but only for a limited number of frames. 

The common bottleneck of these detectors is the limited data bandwidth off the
front-end detector application-specific integrated circuit (ASIC) resulting
from the purely analog design flow. An approach to overcome
this limitation is to move into the digital domain as early as possible. The
transmission rates of digital data can be higher than those of analog data. 
Digital signals also are less prone to corruptions. Error detection and correction 
algorithms can ensure data integrity, and standardized interfaces can be used.
In addition, the digital data manipulation (e.g., compression) can be 
incorporated on the same piece of silicon as the front-end detector. The big 
gain in sub-\SI{100}{\nm} CMOS technology is in the digital 
domain (i.e., clock speed, power, logic density). Using smaller process nodes 
is the most effective path to high logic density. CMOS feature size reduction 
results to first order in a quadratic increase with the feature reduction factor 
\cite{Garcia_Sciveres_2018}. Increasing frame rates can be achieved by using 
multiple high-speed multi-gigabit transceivers on the front-end ASIC. In this 
paper, we present an additional means to increase the effective bandwidth and thus the 
frame rate by using data compression on the front-end detector ASIC. While 
this concept (i.e., bandwidth compression \cite{Ueno:2017hv}) is not new, such
advanced digital concepts have not been exploited for X-ray pixel detectors.
In this paper, we present new strategies to make the most efficient use of the
off-chip bandwidth by utilizing data compression schemes for photon-counting 
and charge-integrating pixel detectors.

\section{Detector Architecture Overview}
\label{sec:DetectorArchitecture}
\begin{figure}[htbp]
\centering 
\includegraphics[width=1\textwidth,origin=c,angle=0]{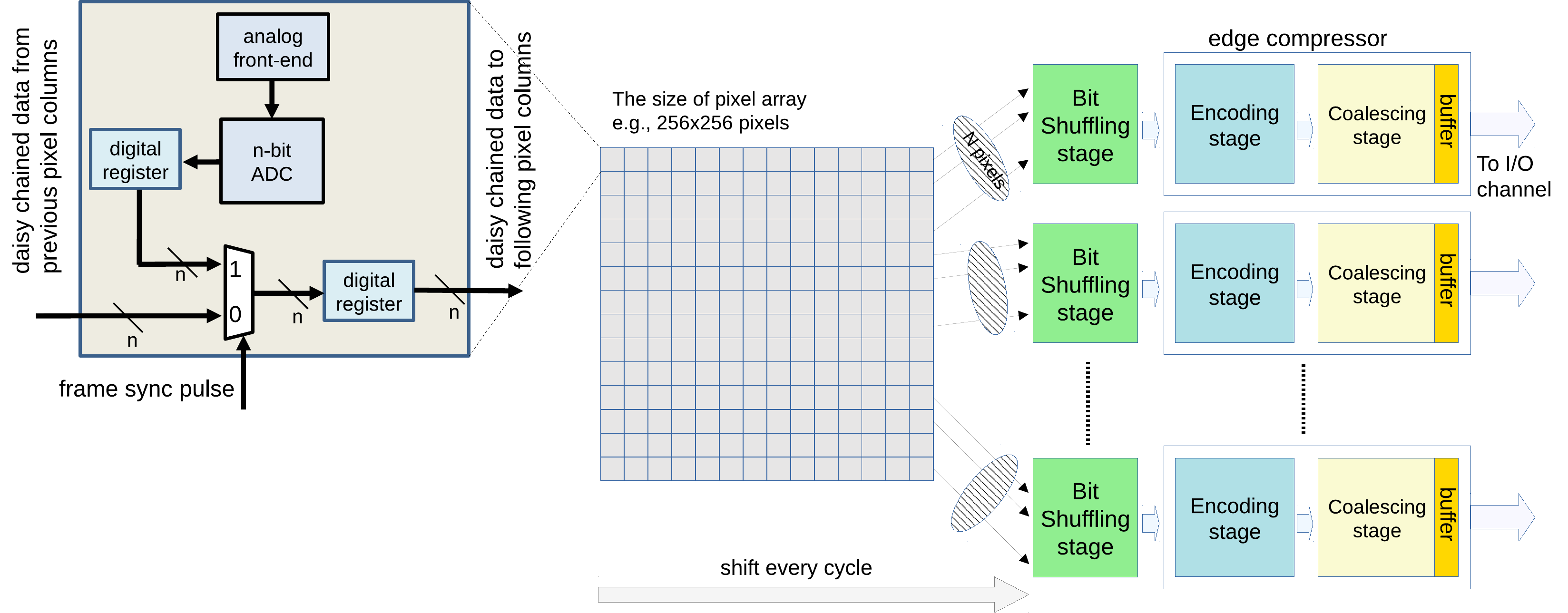}
\caption{Block diagram of the pixel detector ASIC architecture.}
\label{fig:ASIC} 
\end{figure}

To demonstrate our concept, we consider a 
hybrid X-ray pixel array detector that consists of 
a passive sensor array bump-bonded to a mixed-signal
ASIC. In this paper, we present designs in a 
commercial \SI{65}{\nm} mixed-signal CMOS technology and 
aim for a pixel size around $\sim\!\SI{100}{\um}$ per side. 
Underneath each sensor pixel, each pixel of the 
mixed-signal ASIC contains an analog front-end, 
a gain-switching (e.g., adaptive gain or 
autoranging) amplifier, a small analog-to-digital 
converter (ADC), and digital logic. We 
note that the concepts presented here can also be 
applied to other detector topologies where the 
analog and digital logic are located on different 
pieces of silicon that are bonded together. Once 
per frame, the ADC generates a digital value that 
represents the voltage resulting from the 
charge integrated by the front-end during the 
previous frame. The voltage and therefore the 
digital value from the ADC at the end of each frame 
are assumed to be linearly proportional to the 
number of photons detected by the sensor during the 
previous frame. We will utilize digital logic to 
shift the sample data from all ADCs off the device. 
A block diagram of the pixel array detector is shown
in Figure ~\ref{fig:ASIC} that demonstrates the 
different pieces required for a complete 
science-grade detector array on the order of $128 
\times 128$ to $256 \times 256$ pixels. Figure 
~\ref{fig:ASIC} also illustrates the logic inside 
each pixel. Under control of a frame sync pulse, 
once per frame the logic within each pixel loads the
ADC outputs into registers that constitute a long 
shift register spanning an entire pixel column. While 
charge is being integrated by the sensor and 
front-end electronics for the current frame, the 
digital logic is shifting out all digital samples 
from all pixels for the previous frame to the edge 
of the pixel array. The digital logic within each 
pixel can also perform preprocessing functions 
(e.g., denoising and encoding) on the samples prior
to shifting, in order to minimize the number of bits sent to 
the edge. At the edge of the ASIC, compression logic
further reduces the number of bits required to be 
sent through high-speed digital 
transmitters to an off-chip data acquisition system 
(DAQ). A digital memory can be used as an elastic 
store to help smooth out peaks and valleys in the 
data streams being transmitted.

\section{Compression}
\label{sec:Compression}
A detector ASIC has a finite amount of bandwidth to shift sample data from the pixels to the ASIC's edge and 
between the ASIC and external DAQ electronics. Depending on the parameters of a particular experiment, the user of 
the detector may choose either to enable on-chip digital processing and compression, thereby maximizing the frame 
rate, or to disable digital processing and compression and send the uncompressed sample data off-chip 
at a lower frame rate or clock frequency. The ASIC architecture we present will allow fine tuning of how the ASIC's 
bandwidth is used by offering a high level of configurability via software programming. The only limiting factors 
will ultimately be the amount of bandwidth hardwired into the ASIC. 

The following sections discuss three methods for minimizing the amount of data that needs to be transmitted 
off-chip: (1) digitally removing noise from each ADC sample (denoise); (2) digitally encoding ADC outputs to a bit-reduced sequence near the photon Poisson noise level; and (3) streaming
edge compression. Methods 1 and 2 can be implemented within each pixel. Method 3 is implemented at the edge of the 
ASIC, outside of the active sensor area, and therefore is independent of the pixel design and pixel size. All three 
data compression methods work together to make the best use of the available ASIC bandwidth, both within the pixel 
array's shift buses and, more important, in the high-speed transmitters that send the data off-chip. This design  allows the 
detector to achieve the highest possible frame rate within the given bandwidth constraints. The first method of removing noise from each sample helps 
maximize the number of transmitted sample values that equal zero, thus maximizing the effectiveness of the edge 
compression algorithm. The second method reduces the number of bits to be sent by digitally dividing the ADC output 
such that the resulting values efficiently encode the number of photons detected based on the photon Poisson noise level. 
The third method implements a digital compression algorithm that compacts the 
samples with zero values. Since all three  data reduction 
techniques are digital, they lend themselves to a high degree of programmability.

\subsection{Compression in the pixel}
\subsubsection{Digital denoising}
\label{sec:denoising}
In order to maximize the number of zero-values pixels that are sent to the 
edge compressor, the raw ADC output 
should have all samples that represent only noise zeroed out. This can be achieved in the
digital domain by determining a digital noise floor value for each pixel. 
This is the maximum digital value on the ADC outputs when no photons are detected. This 
value is determined during a calibration sequence by exposing the detector to darkness, 
integrating for a specified interval, and examining the ADC outputs. The resulting noise 
floor is programmed into a control register within each
pixel. Any ADC sample that is below the noise floor threshold is forced to 0 
by logic. Digital denoising is  necessary only for charge-integrated detectors. 
Since this operation discards the information, it should be considered lossy compression. For example, digital denoising of charge-integrated detectors discards information related to charge-shared events that could be used for subpixel resolution \cite{gotthard} or energy resolution \cite{SeidlerColorSubpixel}.  However, digital denoising of charge-integrated detectors is essentially a photon-counting detector, and thus we do not expect significant issues with scientific data quality.

\subsubsection{Digitally encoding near the photon Poisson noise}
\label{sec:in-pixel-encoding}
If the detector simply transmits the raw ADC output value each frame, 
many of the transmitted bits are superfluous since the available resolution of the ADC may already exceed the Poisson noise of the incoming photon flux. By performing some simple digital operations on the ADC
output, the number of bits per sample can be reduced. This compression scheme is lossy in nature and thus needs to validated on scientific data. In a companion paper, we used the compression scheme described in this section and  showed that  encoding the raw 14-bit photon-counting detector output into only 8 or 9 bits has a negligible effect on the ptychographic image reconstructions \cite{Panpan}. This section describes an in-pixel encoding scheme where the ADC value is digitally divided to create an encoded sequence with reduced bit count. Two examples are presented. Example 1 is a simpler encoding that applies to photon-counting detectors and encodes the photon count directly with no intermediate ADC value. Example 2 is more complex and applies to charge-integrating detectors with gain-switching front-ends. Example 2 also includes an explanation of how the method can be implemented with simple digital logic. 

\begin{figure}[htbp]
\centering 
\includegraphics[width=1\textwidth,origin=c,angle=0,trim=0 0 0 25,clip]{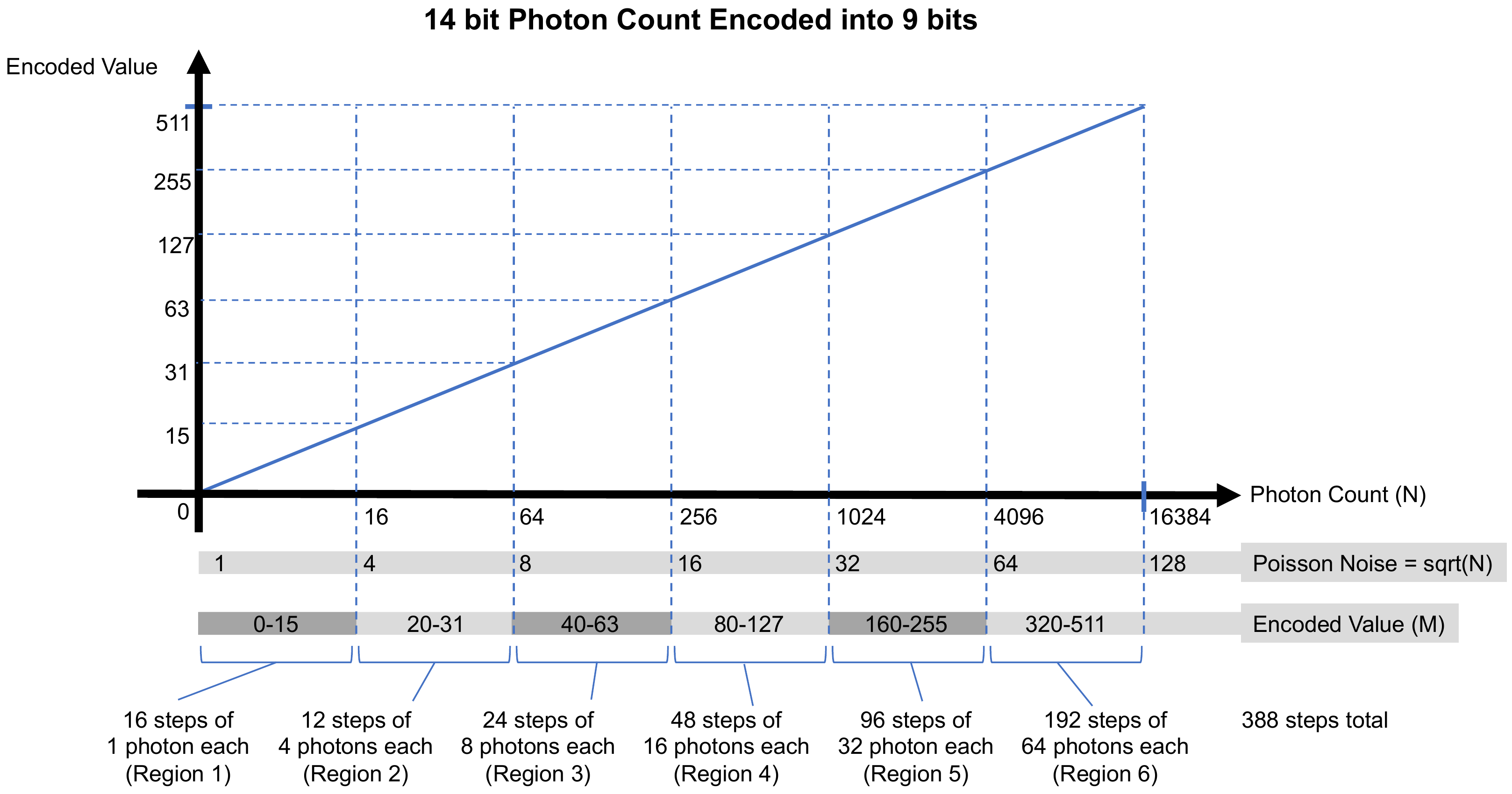}
\caption{Encoding for a 14-bit photon-counting front-end detector into 9 bits. The photon count range is divided into 6 regions. At the start of each region, the Poisson noise is shown as well as the encoding in each range.}
\label{fig:encoding-counting} 
\end{figure}

Example 1, shown in Figure ~\ref{fig:encoding-counting}, demonstrates how
to encode a photon count N into fewer values. This example shows a 14-bit 
photon count (0--16384) encoded into 9 bits (0--511). As the photon count N 
increases, so
too does the Poisson noise as $\sqrt{N}$. Therefore, as the photon count 
increases, each encoded value can represent a larger number of photons per 
step. In the figure, the entire photon count range is divided into 6 regions; details are shown in Table ~\ref{table:photon_count_table}. In this example, we have chosen the step size exactly equal to the Poisson noise (i.e., $\sqrt{2}$ increase in total noise) at the lower bound of the encoding range.  At least for ptychography data \cite{Panpan}, this level of lossy noise has negligible effect on the scientific data quality; we have yet to study its effects on other experimental techniques. 

\begin{table}[]
\centering
\begin{tabular}{|c|c|c|c|}
\hline
Region & \begin{tabular}[c]{@{}c@{}}Photon Count \\ Range\end{tabular} & \begin{tabular}[c]{@{}c@{}}Number of \\ Photons per \\ Encoded Step\end{tabular} & \begin{tabular}[c]{@{}c@{}}Number of \\ Encoded Steps\end{tabular} \\ \hline
1 & 0-15 & 1 & 16 \\ \hline
2 & 16-63 & 4 & 12 \\ \hline
3 & 64-255 & 8 & 24 \\ \hline
4 & 256-1023 & 16 & 48 \\ \hline
5 & 1024-4095 & 32 & 96 \\ \hline
6 & 4096-16383 & 64 & 192 \\ \hline
\end{tabular}
\caption{Number of encoded steps per photon count for the photon-counting front-end example shown in Figure ~\ref{fig:encoding-counting}.}
\label{table:photon_count_table} 
\end{table}

In region 1 of Figure ~\ref{fig:encoding-counting} and Table ~\ref{table:photon_count_table}, the encoded values each represent 1 photon per step. These values cannot easily be encoded since detection of single photons is an important requirement for a detector. In region 2, however, the Poisson noise equals 4, so each encoded value can represent 4 photons per step, and the entire 48-photon region can be encoded into 12 values.

In region 6, for a detector to report individual photon counts of 4,096--16,383 is wasteful of bandwidth since the Poisson noise is 64 or greater in this region. By encoding the photon counts such that each encoded value represents a range of 64 photon counts, the number of values is reduced from 12,288 to just 192 with essentially no loss of information. The greatest efficiency in encoding comes at higher photon counts where the Poisson noise allows each unique encoded value to represent a larger range of photon counts. 

As will be explained in Example 2, the boundaries of each region are chosen to be powers of 2 to simplify the encoding logic. An artifact of this is that there are gaps in the encoding sequence since not all encoded values between 0 and 511 are needed.

\begin{figure}[htbp]
\centering 
\includegraphics[width=1\textwidth,origin=c,angle=0,trim=0 0 0 25 ,clip]{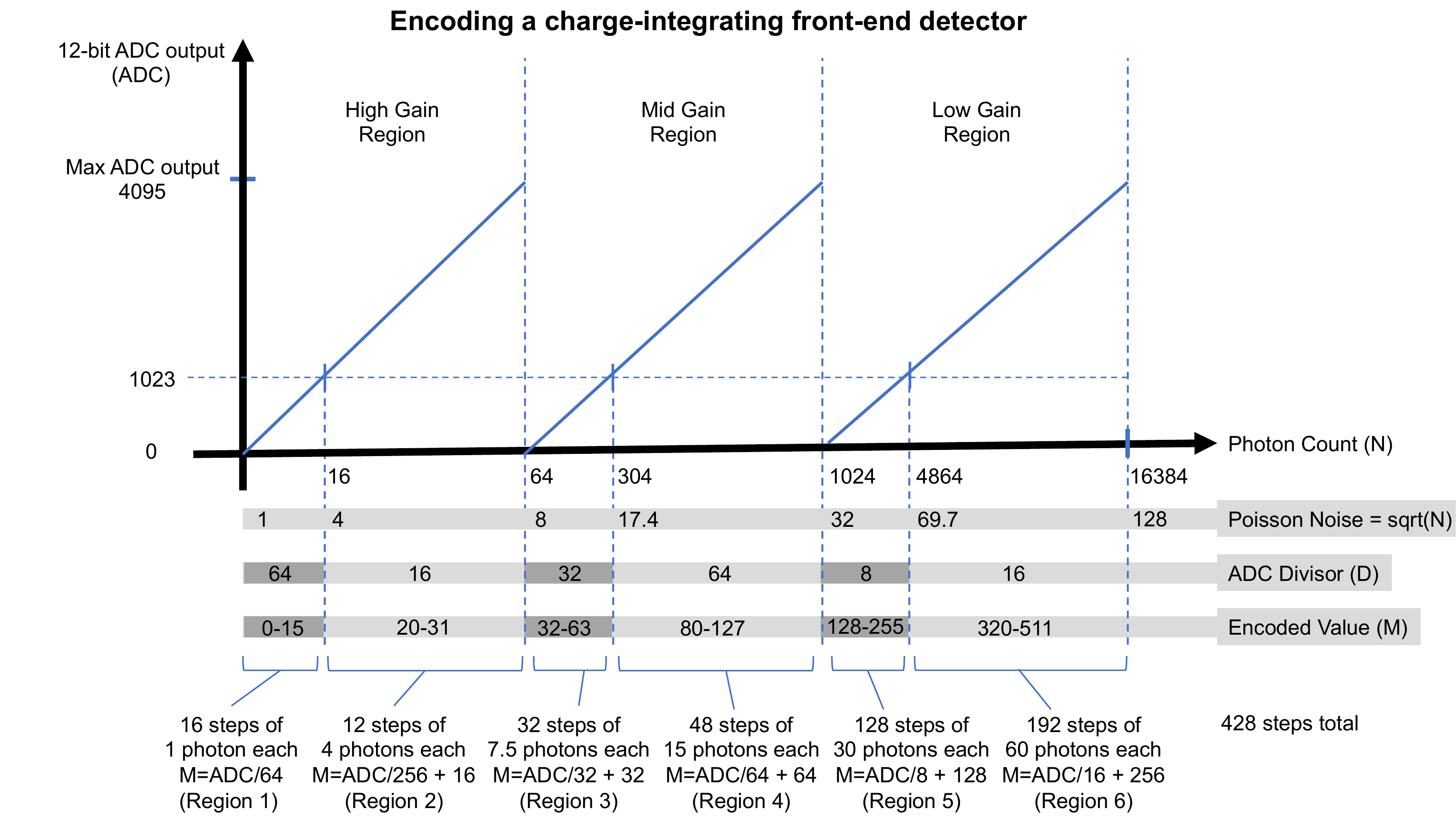}
\caption{Charge-integrating encoding showing the number of photons as a function of a 12-bit ADC digital output 
for a three-stage gain front-end. The Poisson noise and the encoding of ADC outputs into bit 
reduced sequence are shown. The required number of transmitted bits per pixel is reduced from 14 bits to 9 bits.}
\label{fig:encoding-3gains} 
\end{figure}

Example 2 is shown in Figure ~\ref{fig:encoding-3gains} and assumes a pixel design with a charge-integrating front-end and automatic gain switching. Here we present an idealized analog front-end for the sake of explanation. However,  a number of issues  need to be addressed in order to realize this scheme in practice, including operating at different photon energies, nonconstant gain, nonzero offsets, and matching of the entire analog swing to ADC range. We expect innovations in on-chip calibration of charge-integrating pixels detectors to resolve some of these issues in the future. 

This example assumes 3 analog gain regions: high gain, medium gain, and low gain, as is typical in existing detectors. The three gain regions are each subdivided into two subregions, making 6 regions in total, as labeled in the figure. This example shows how the encoding algorithm can be implemented in a few logic gates, making it practical to easily incorporate within each pixel and thereby reducing the number of bits that need to be sent from each pixel to each frame. Each gain region is associated with a specific detected photon count range. Within each region there is assumed to be a linear relationship between the number of photons detected and the ADC output. Typically detectors might transmit the 12-bit ADC value to the DAQ each frame, along with 2 gain bits indicating the gain region, for a total of 14 bits (ADC+gain) per pixel per frame.

In the high-gain region one can easily see that reporting a 14-bit digital value corresponding to an ADC output of 0--4095 but representing only 0--63 detected photons is wasteful since the photon count could be encoded into only 6 bits. This can be achieved by dividing the ADC output by 64 and sending the resulting value. The number of encoded values can be even further reduced by taking advantage of the algorithm described in Figure ~\ref{fig:encoding-counting} and increasing the photon count per encoded value as the Poisson noise increases. Figure~\ref{fig:encoding-3gains} shows encoding photon counts of 16--63 into 12 values representing 4 photons per step. This reduces the original 0--63 photon counts to just 28=16+12 encoded values.

As described earlier, the encoding efficiency increases as the photon count increases. In the low-gain region our example detects 1,024 to 16,383 photons represented by ADC output values of 0--4095. Each ADC step represents less than 4 photons (4096/(16384-1024). However, the Poisson noise is >32 in the low gain region. The ADC resolution exceeds what the Poisson noise of the incident beam supports. The reported value could be reduced from 4,096 unique ADC values down to 480=(16384 - 1024)/32 encoded values, and the digitization noise would be at least as good as the Poisson noise in all cases. Additionally, if the encoding logic divides the low-gain region into two subregions such that more photon counts are encoded into each step after the ADC reaches 1/4 of its full range, then the total number of encoded values is reduced even further to 320=128+192.

The threshold values chosen in Figure ~\ref{fig:encoding-3gains} have been carefully selected so that the encoding algorithm can easily be implemented in digital logic with very few gates. This allows the circuit to be added to each pixel since it will occupy very little space and consume very little power. To encode the ADC+gain bits, the digital logic performs the following three steps:
\begin{enumerate}
\item Determine which of 6 subregions the ADC is reporting for. In logic, this means examining the 2-bit gain value to determine 1 of 3 gain regions and examining the 2 most-significant bits (MSBs) of the ADC output to determine whether the value is above or below 1,023.

\item Divide the 12-bit ADC output by the divisor D. In digital logic dividing by a power of 2 is just a simple binary shift. For example, to divide the 12-bit ADC output representing values 0--1023 by a divisor of 64 means removing the 6 least significant bits and padding the top 6 MSBs with zeros. No digital divider circuit is actually needed.

\item Add the divided value from step 2 with an offset value to create a monotonically increasing encoding. In our example, all of the added offsets are powers of 2. For example, region 6 requires M=ADC/16 + 256. Dividing by 16 means right shifting the 12-bit ADC value by 4 bit positions. Adding 256 means setting the resulting 9\ts{th} bit position to 1. No digital adder circuit is actually needed.
\end{enumerate}

\begin{figure}[htbp]
\centering 
\includegraphics[width=1\textwidth,origin=c,angle=0]{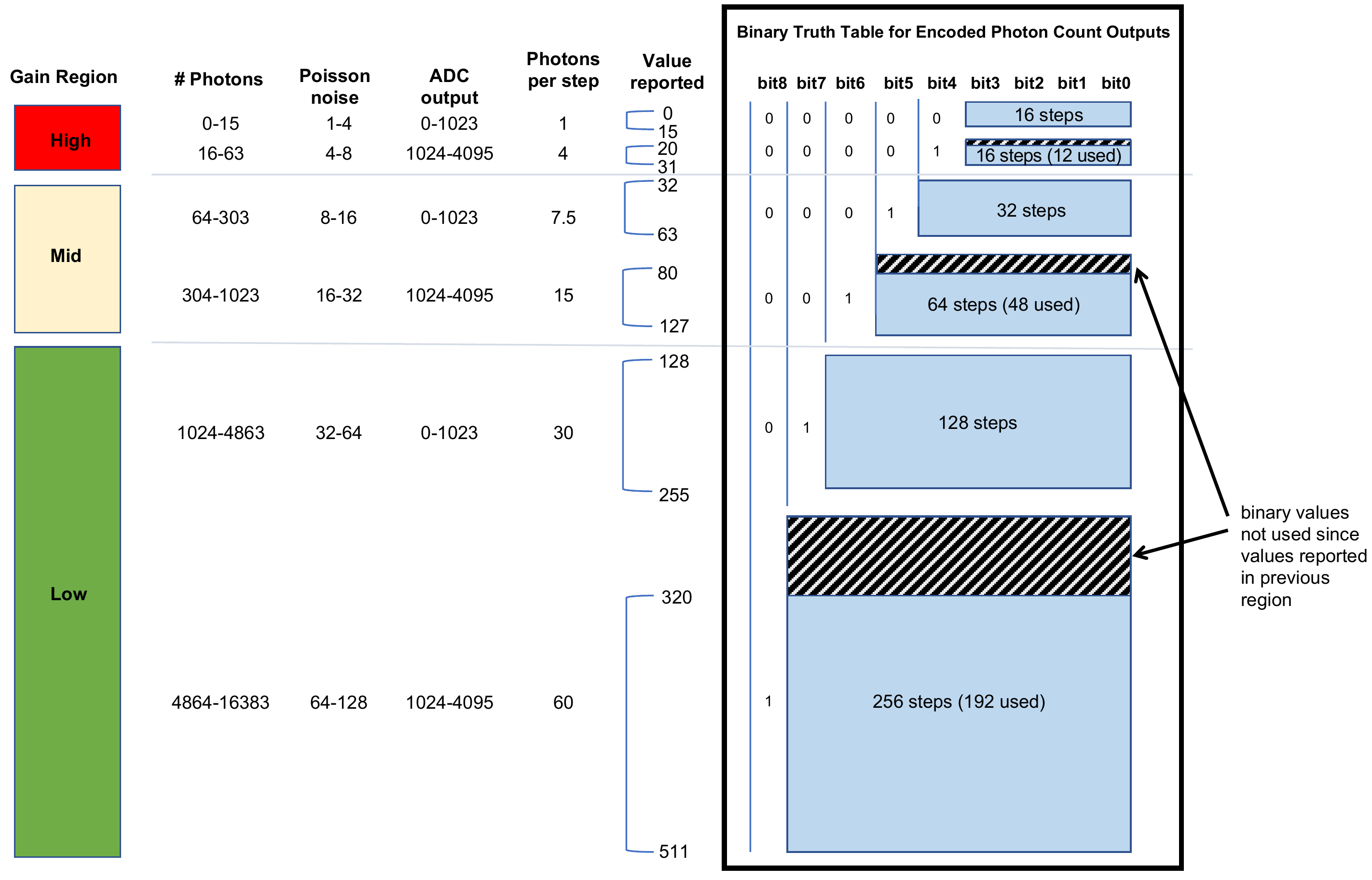}
\caption{Truth table of the encoding of ADC outputs into bit-reduced sequence. }
\label{fig:encoding-truth-table} 
\end{figure}

Figure \ref{fig:encoding-truth-table} shows the encoding in truth table format. Gaps in the final encoded sequence indicate where there are unused values. These are because the offset additions are chosen to be powers of 2 to make the digital logic simpler. These unused values could be removed in logic such that the final sequence had no gaps, but this would come at the expense of more complicated digital logic, and it ultimately would not change the detector performance. 

\begin{figure}[htbp]
\centering 
\includegraphics[width=1\textwidth,origin=c,angle=0]{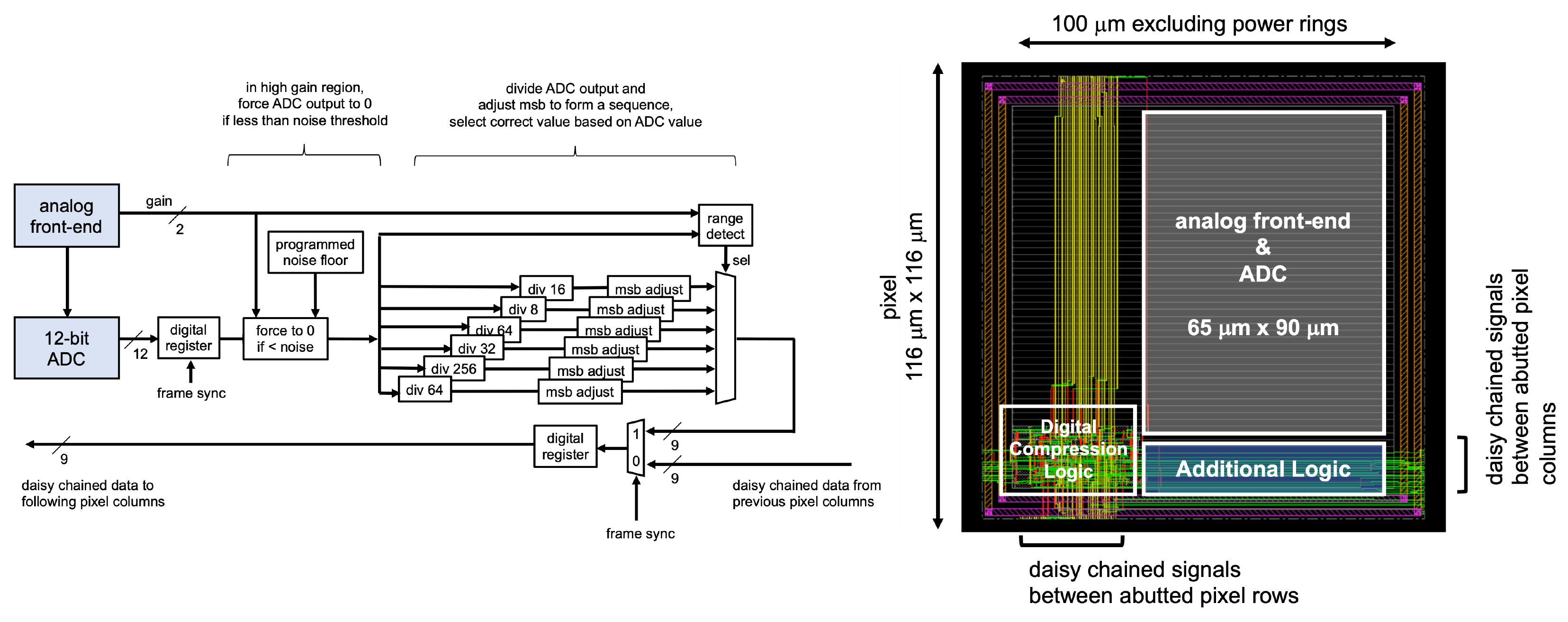}
\caption{Block diagram and physical layout of digital logic for in-pixel encoding algorithm.}
\label{fig:encoding-block-layout} 
\end{figure}

Figure \ref{fig:encoding-block-layout} shows a block diagram of the logic that implements this encoding algorithm that has been coded in RTL, synthesized, and routed in \SI{65}{\nm} standard cell CMOS technology. At approximately \SI{25}{\um} per side this encoding logic is small enough to be incorporated into each pixel. Alternatively, since it is small and will run fast, it can be placed at the edge of the ASIC and shared across pixel data streaming from the column throughout the frame. The trade-off in placing the encoding logic within each pixel or at the edge of a pixel column is more digital logic overall with fewer shifts bus wires versus less digital logic but more wires. An example layout of a pixel containing this logic is shown. 

\begin{figure}[htbp]
\centering 
\includegraphics[width=1\textwidth,origin=c,angle=0]{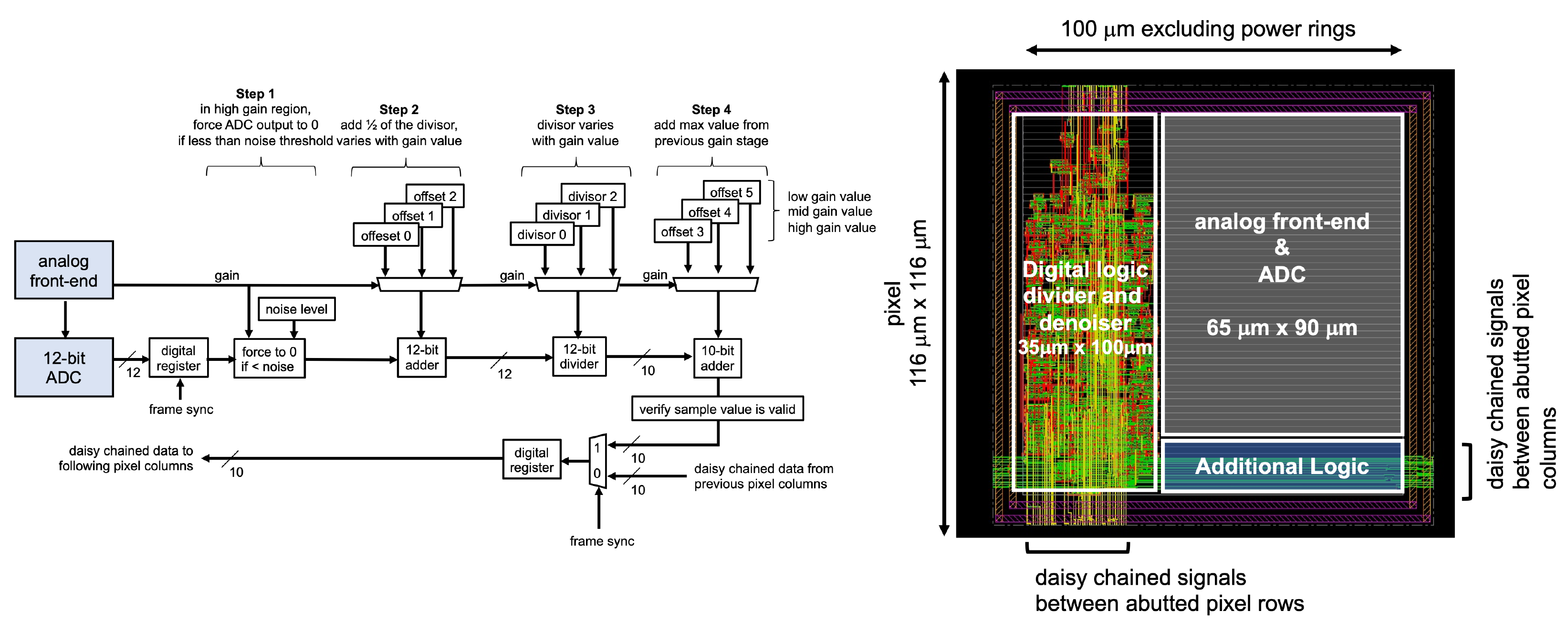}
\caption{Block diagram and physical layout of digital logic for a pixel including 12-bit programmable divider and denoiser.}
\label{fig:divider-block-layout} 
\end{figure}

For comparison, a different circuit is shown in Figure \ref{fig:divider-block-layout} that incorporates a 12-bit programmable integer divider. This design was coded in RTL and implemented. This circuit allows more flexibility in the algorithm since divisors and threshold values can be selected after the ASIC is designed, unlike the fixed power-of-2 thresholds and divisions in the algorithm presented above. The programmable divider approach comes at the expense of about $4\times$ as much digital logic and more than $2\times$ the power of the fixed division approach, but it can provide a similar compression and reduction of the number of bits that need to be transmitted. The digital logic of either approach can support many design features, such as disabling of compression and checks to ensure that valid sample data are being transmitted.

An alternative implementation of the encoding algorithm could use an SRAM programmed as a lookup table. The ADC output and gain bits could be applied to the read address inputs of the SRAM, and the output read data would be the encoded value. 
A hybrid approach can use logic to reduce the required address space by assigning groups of larger photon count ranges to common read addresses, minimizing the size of the SRAM needed. This approach allows the encoded values to be programmed after device fabrication. It can also allow the encoded values to more closely follow the photon Poisson noise curve. In \SI{65}{\nm} technology the physical size of the memory required would be too large to place one in each pixel, but it would be possible to place look-up tables at the ends of the pixel columns and time-share them across pixels.

The in-pixel encoding scheme described in this section differs from traditional data compression schemes in two ways. First, this encoding scheme produces a fixed-bit reduction in percentage regardless of the nature of the data, which simplifies the digital transmission logic. Second, there is no need to decompress the data off the chip since we have simply represented the original data in a bit-reduced manner.

\subsection{Compression at the detector ASIC edge}
\label{sec:edge-compression}

The in-pixel encoding scheme described above makes no assumptions about the nature of the data measured by the detector. However, we can exploit redundancies in the data to further compress the data. In this section, we describe a lossless compressor located at the ASIC edge that exploits the most common redundancy in X-ray data---the zeros. The aim of lossless compression is to reduce data by identifying and eliminating statistical redundancy. The characteristics of input data affect both the complexity of the design and the compression ratio. In general, complex designs require significant effort to validate and test in an ASIC. Our goal is to identify a simple design that fulfills two requirements. 
First, it must yield reasonably high compression ratios for typical X-ray datasets.  Second, operation must be stall-free; it must process input pixel data shifted from the pixel array every single clock cycle without causing any stall clock cycle or dropping any data. The design should also be simple and have a minimal resource footprint on the pixel detector ASIC with no dependency on any external IP libraries, which also reduces the ASIC validation and testing efforts dramatically.

\subsubsection{Analysis of four typical X-ray datasets}
\begin{figure}[htbp]
\centering 
\includegraphics[width=1\textwidth,origin=c,angle=0]{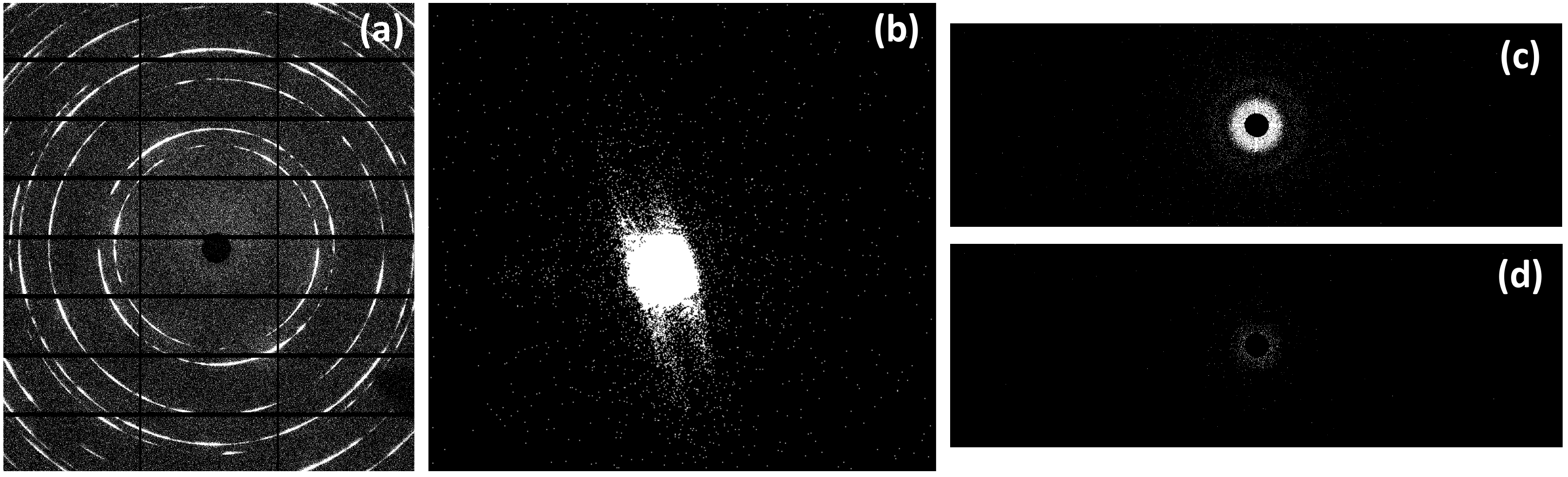}
\caption{Four representative images for the (a) high-energy XRD, (b) ptychography, (c) XPCS concentrated, and (d) XPCS dilute datasets. Pixel values equal to zero are displayed in black and $\geq\!1$ in white. Note that the ptychography dataset taken with the Eiger 500K has been cropped to $558 \times 514$ pixels.}
\label{fig:x-ray-data} 
\end{figure}

We have analyzed four typical X-ray datasets from a third-generation storage ring light source to guide the design of the edge compressor. The first dataset is typical of high-energy X-ray diffraction (XRD) experiments of polycrystalline materials; the data consist of 300 images from a time-resolved additive manufacturing experiment at the Advanced Photon Source, APS 1-ID-E, taken with a Dectris Pilatus3 X CdTe 2M detector. The second dataset is typical of a ptychography  experiment; the data consist of 1,737 images from a ptychography experiment at APS 2-ID-D taken with the Dectris Eiger X 500K detector. The third and fourth datasets consist of typical X-ray photon correlation spectroscopy (XPCS) experiments with concentrated and dilute samples of polymer spheres diffusing in glycerol; the data consist of 1,000 images taken at APS 8-ID-I with the X-Spectrum Lambda 750K detector. Figure \ref{fig:x-ray-data} shows a single image from each dataset. All three datasets were taken with photon-counting detectors that we use to emulate a charge-integrating detector and have been digitally denoised as described in Section~\ref{sec:denoising}. As one can see, the majority of pixels in all the images are zero. Table~\ref{table:zero} includes a statistical analysis of the high-energy XRD, ptychography, XPCS concentrated, and XPCS dilute datasets on the zero-valued pixels and a compression ratio that gzip offline compression can achieve as a reference. The percentage of zero-valued pixels is calculated every frame. The mean, standard deviation, minimum, and maximum are calculated from the percentage of zero-valued pixels of all frames. As the brightness of upgraded storage ring light sources increases, the percentage of zero-valued pixels may decrease. However, new X-ray sources will allow scientists to study more weakly scattering samples with higher frame rate detectors, so the percentage of zero-valued pixels may stay the same or decrease. The four X-ray datasets have a wide range of zero-valued pixels that enable us to explore the limits of our streaming edge compressor.  The compression ratio of the gzip offline compressor, which employs Lempel-Ziv coding (LZ77), is obtained from the ratio between the original file size and the compressed file size of a dataset.

\begin{table}[]
\begin{center}
\begin{tabular}{|l|c|c|c|c||c|}
\hline
\multicolumn{1}{|c|}{\textbf{Dataset}} &
  \begin{tabular}[c]{@{}c@{}}Mean\\ \% \end{tabular} &
  \begin{tabular}[c]{@{}c@{}}Std Dev\\ \% \end{tabular} &
  \begin{tabular}[c]{@{}c@{}}Minimum\\ \% \end{tabular} &
  \begin{tabular}[c]{@{}c@{}}Maximum\\ \% \end{tabular} &
  \begin{tabular}[c]{@{}c@{}}offline gzip\\compression ratio\end{tabular} \\ \hline
High-energy XRD   & 83.42 & 2.49 & 68.96 & 85.03  & 19 \\ \hline
Ptychography      & 97.42 & 0.23 & 96.2 & 97.77  & 70 \\ \hline
XPCS concentrated & 98.62 & 0.03 & 98.02 & 98.68  & 67 \\ \hline
XPCS dilute       & 99.9 & 0.01 & 99.8 & 99.91  & 351 \\ \hline
\end{tabular}
\caption{Statistics of zero-valued pixels per frame and offline software-based compressor results.}
\label{table:zero} 
\end{center}
\end{table}

\subsubsection{Design encoding scheme}
Since the majority of pixels in all the datasets are zero-valued pixels, a
natural thought is to employ run-length encoding (RLE) \cite{RLE},
which leverages the consecutiveness of the same value in the input sequence.
As a concrete example, an input sequence [2, 0, 0, 0, 0, 0, 0, 0] would be encoded into [1, 2, 7, 0] using one pixel storage (e.g., 9 bits) for the number of the occurrences of a pixel value and another pixel storage for the pixel value itself. In this example, the compression ratio is $2$.
Theoretically, as the length of the input sequence increases, the compression ratio increases.  According to the percentage of zero-valued pixels in Table~\ref{table:zero}, approximately 5 out of 6 pixels,
38 out of 39 pixels, 71 out of 72 pixels, and 999 out of 1,000 pixels are a zero-valued pixel in high-energy XRD, ptychography, XPCS concentrated, and XPCS dilute, respectively, thus allowing us to estimate the length of continuous zero-valued pixels. If 5 out of 6 pixels are continuous (the best case), four pixels are needed to encode 6 pixels, which yields $1.5\times$ in the compression ratio on high-energy XRD. Applying the same calculation to XPCS dilute, which has the highest percentage of zero-valued pixels, the theoretically possible maximum compression ratio would be $250\times$.
However, zero values typically are distributed nonuniformly  through an X-ray image, which leads to higher variability in compression ratios. In the worst case for RLE, the output size is twice  the input size. Such higher variability in the encoded data size imposes a significant challenge to hardware design when it comes to I/O handling and can possibly be a source of stall. Thus we need an encoding scheme whose encoded data have little variability in the encoded data size.

\begin{figure}[htbp]
\centering
\includegraphics[width=.8\textwidth,origin=c,angle=0]{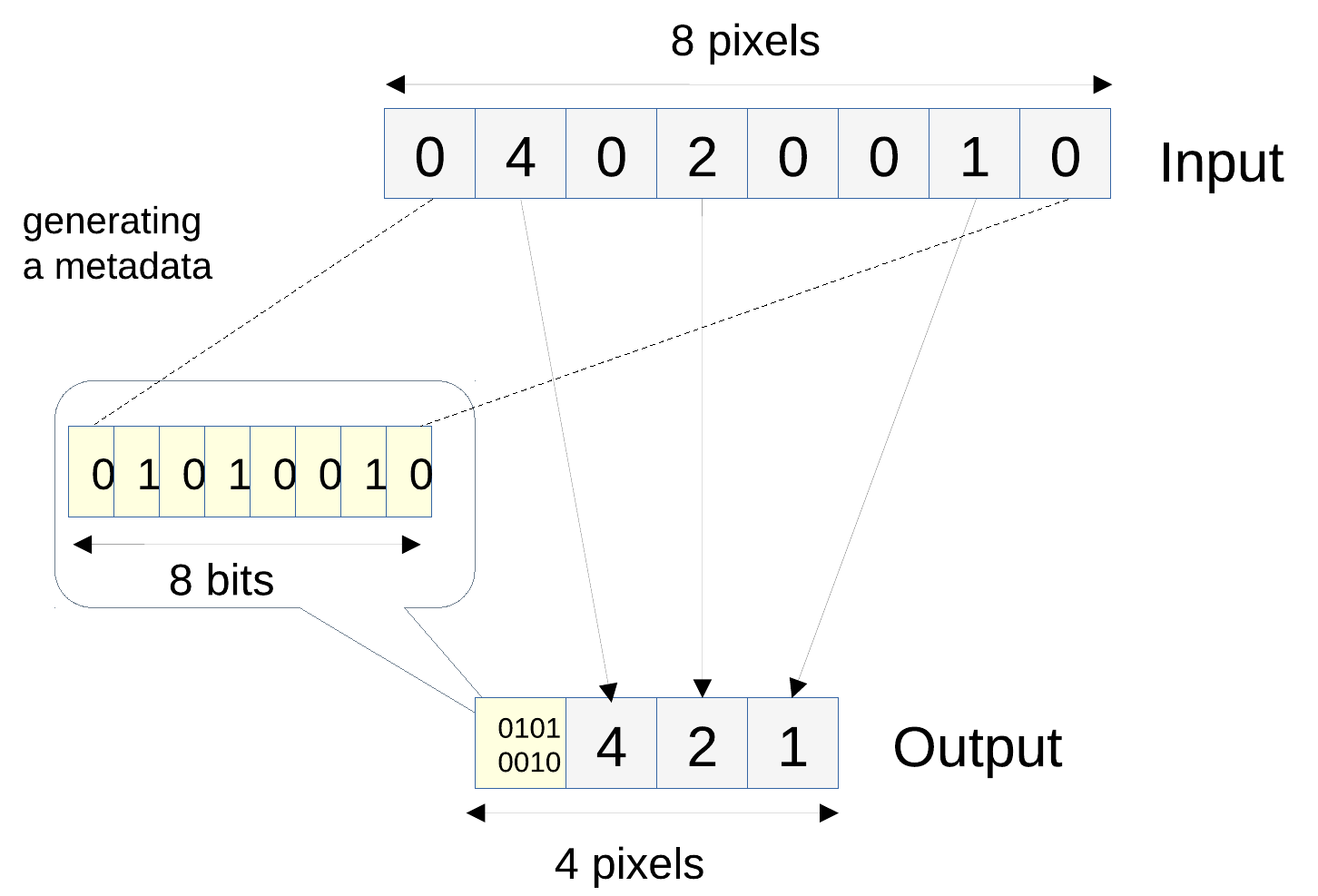}
\caption{Zeromask encoding scheme.}
\label{fig:zme} 
\end{figure}

In this paper, we propose a new compression encoding scheme named ``zeromask'' (ZM) that compresses data by leveraging zero pixels. Figure~\ref{fig:zme} depicts how ZM encodes input data with an example. The output of ZM consists of metadata and non-zero-valued pixels, where the metadata are a sequence of bits that preserves the original position of zero and nonzero pixels in the input segment. Unlike RLE, this approach does not require zero pixels to be contiguous and is thus effective with a more random input as long as it contains zero pixels. To simplify the logic, we repurpose a pixel storage for the metadata.  With 9-bit pixel storage, it can hold a metadata for up to 9 pixels. Since the size of pixel arrays is generally bound to the power of 2 (e.g., 256x256), a reasonable choice for the number of input pixels (N) to each ZM compressor logic would be 8 instead of 9.
The maximum length of the encoded data will be $N + 1$ when all pixels are nonzero; this yields the worst compression ratio ($0.88\times$ with N=8), which outperforms RLE with $0.5\times$ compression ratio; the minimum length of encoded data being one, which  includes only metadata,  yields the best compression ratio ($8\times$ with N=8).

\begin{figure}[htbp]
\centering
\includegraphics[width=.6\textwidth,origin=c,angle=0]{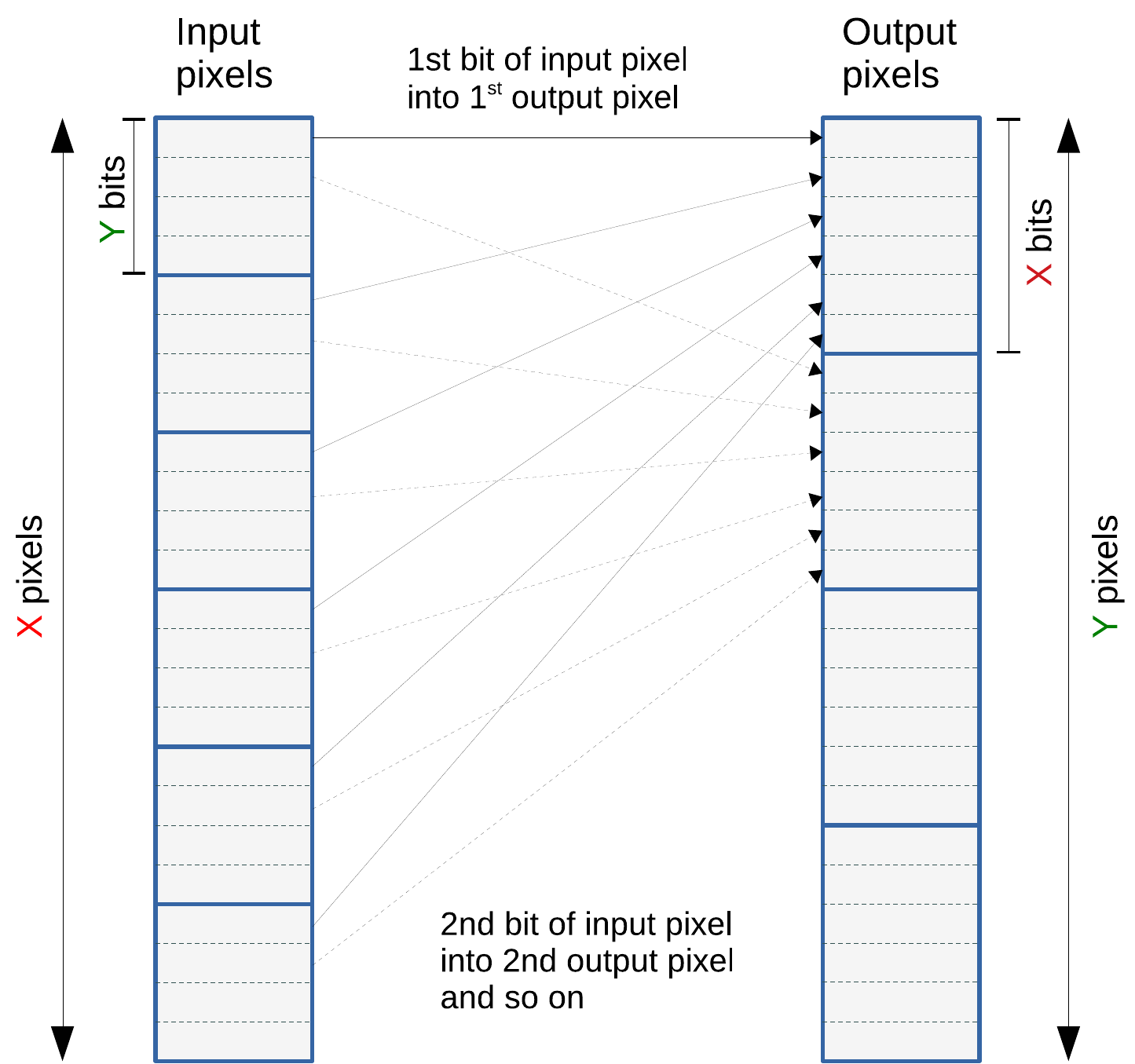}
\caption{Bit-shuffling scheme.}
\label{fig:bitshuffle} 
\end{figure}

An obvious restriction of ZM is that the maximum compression ratio is limited by the number of input pixels and the number of bits per pixel, which is equal to the number of bits per metadata. To relax this restriction, we employ a bit-shuffling operation that can effectively increase the input size by shuffling bits (Figure~\ref{fig:bitshuffle}); the operation  greatly resembles \textit{BitShuffle}~\cite{Bitshuffle}, a software implementation of a bit-level transpose operation. We denote the original input by $p(i)_j$, where i is the index of the pixel and j is j\ts{th} bit. The bit-shuffling operation simply converts $p(i)_j$ to $q(j)_i$. This operation resembles a matrix transpose and is reversible and thus decodable. 
With a 9-bit pixel, it converts 16 input pixels to 9 16-bit integer blocks.
The bit-shuffling operation is simply a set of wires between the input bits and the output bits in a correct order and requires no logic circuit in ASIC; hence it is inexpensive to implement and verify.
After the bit shuffling, we can simply use ZM to compress data. No redesign of ZM is needed; only the parameters need to be changed. The bit shuffling is effective for the X-ray datasets, particularly high-energy XRD, where a majority of nonzero pixels use only the lower bits such as the pixel value 1--4 or three least-significant bits. For example, a sequence of 9-bit input pixels [1, 2, 3, 1, 0, 2, 3, 1] is converted to a sequence of 9 integer blocks [179, 68, 0, 0, 0, 0, 0, 0, 0], where each integer block is 8 bits.\footnote{The decimal number 179 is 10110011 in binary, and 68 is 01000100.}

To evaluate different encoding schemes, we also developed a software tool\footnote{The tool is written in Scala so that we can reuse codes in Chisel testbenches to verify the outputs from simulated hardware.} that reads actual X-ray datasets from a file and compresses input data by using different encoding schemes. Each dataset frame is divided by eight or sixteen rows, depending on the input size to the encoder.  For example, with the XPCS datasets ($1556\times512$ pixels), each frame is broken into 64 chunks with eight input pixels and 32 chunks with sixteen input pixels, where each chunk has 1,556 columns. Columns in each chunk are fed into encoders one by one,  emulating the shift operation of the pixel array. It iterates the same process for all chunks in each frame. We evaluate the compression ratio of all frames in all four X-ray datasets. The compression ratio is calculated by dividing the total number of the input pixels per frame by the total number of the encoded pixels.  Tables \ref{table:crxrd}, \ref{table:crpty}, \ref{table:crxpcsc},  and \ref{table:crxpcsd} show the compression ratios obtained on the four datasets with different encoding schemes as well as statistics of the variation across all the images in each dataset. The bit width of input pixels is 9 bits.
Of  the three encoding schemes (run-length, zeromask, and shuffled zeromask),
the shuffled zeromask scheme achieves the highest compression for all four datasets and the lowest relative standard deviation. Moreover,  the zeromask scheme achieves the second best compression ratio with a comparatively lower relative standard deviation. 
Earlier in Table~\ref{table:zero} the gzip compression ratios for the four datasets were presented and are significantly larger than shuffled zeromask. However, that gzip compression was performed on the entire dataset. If gzip is used over a smaller dataset, its performance is severely degraded. We have also evaluated the compression of gzip on smaller input sizes. We compressed 8 or 16 bytes input patterns (all zero or all different) using a Python gzip library, which internally uses the standard zlib library. The compression results shown in Table~\ref{table:smallgzip}  clearly indicate that gzip is inefficient for smaller input sizes.


\begin{table}[]
\begin{center}
\begin{tabular}{|l|c|c|c|c|c|c|}
\hline
\multicolumn{1}{|c|}{\textbf{Encoding}} &
  \begin{tabular}[c]{@{}c@{}}Input\\pixels\end{tabular} &
  \begin{tabular}[c]{@{}c@{}}Mean \\ CR\end{tabular} &
  \begin{tabular}[c]{@{}c@{}}Std Dev\\ CR\end{tabular} &
  \begin{tabular}[c]{@{}c@{}}Minimum\\ CR\end{tabular} &
  \begin{tabular}[c]{@{}c@{}}Maximum\\ CR\end{tabular} \\ \hline
Run-length & 8 & 1.4 & 0.08 & 1.0 & 1.5 \\ \hline
Run-length & 16 &  1.6 & 0.11 & 1.0 & 1.7 \\ \hline
Zeromask & 8 &  3.5 & 0.23 & 2.3 & 3.6 \\ \hline
Shuffled zeromask & 16 & 4.2 & 0.13 & 3.5 & 4.3 \\ \hline
\end{tabular}
\caption{Compression ratios of various encoding schemes---high-energy XRD.}
\label{table:crxrd} 
\end{center}
\end{table}

\begin{table}[]
\begin{center}
\begin{tabular}{|l|c|c|c|c|c|c|}
\hline
\multicolumn{1}{|c|}{\textbf{Encoding}} &
  \begin{tabular}[c]{@{}c@{}}Input\\pixels\end{tabular} &
  \begin{tabular}[c]{@{}c@{}}Mean \\ CR\end{tabular} &
  \begin{tabular}[c]{@{}c@{}}Std Dev\\ CR\end{tabular} &
  \begin{tabular}[c]{@{}c@{}}Minimum\\ CR\end{tabular} &
  \begin{tabular}[c]{@{}c@{}}Maximum\\ CR\end{tabular} \\ \hline
Run-length & 8 &  3.3 & 0.05 & 3.0 & 3.3 \\ \hline
Run-length & 16 &  5.4 & 0.14 & 4.7 & 5.6 \\ \hline
Zeromask & 8 &  6.6 & 0.10 & 6.1 & 6.7 \\ \hline
Shuffled zeromask & 16  & 7.3 & 0.07 & 6.9 & 7.4 \\ \hline
\end{tabular}
\caption{Compression ratios of various encoding schemes---ptychography.}
\label{table:crpty} 
\end{center}
\end{table}

\begin{table}[]
\begin{center}
\begin{tabular}{|l|c|c|c|c|c|c|}
\hline
\multicolumn{1}{|c|}{\textbf{Encoding}} &
  \begin{tabular}[c]{@{}c@{}}Input\\pixels\end{tabular} &
  \begin{tabular}[c]{@{}c@{}}Mean \\ CR\end{tabular} &
  \begin{tabular}[c]{@{}c@{}}Std Dev\\ CR\end{tabular} &
  \begin{tabular}[c]{@{}c@{}}Minimum\\ CR\end{tabular} &
  \begin{tabular}[c]{@{}c@{}}Maximum\\ CR\end{tabular} \\ \hline
Run-length & 8 & 3.6 & 0.01 & 3.4 & 3.6 \\ \hline
Run-length & 16 & 6.4 & 0.03 & 5.9 & 6.4 \\ \hline
Zeromask & 8 & 7.2 & 0.02 & 6.9 & 7.2 \\ \hline
Shuffled zeromask & 16 & 8.3 & 0.02 & 8.0 & 8.4 \\ \hline
\end{tabular}
\caption{Compression ratios of various encoding schemes---XPCS concentrated.}
\label{table:crxpcsc} 
\end{center}
\end{table}

\begin{table}[]
\begin{center}
\begin{tabular}{|l|c|c|c|c|c|c|}
\hline
\multicolumn{1}{|c|}{\textbf{Encoding}} &
  \begin{tabular}[c]{@{}c@{}}Input\\pixels\end{tabular} &
  \begin{tabular}[c]{@{}c@{}}Mean \\ CR\end{tabular} &
  \begin{tabular}[c]{@{}c@{}}Std Dev\\ CR\end{tabular} &
  \begin{tabular}[c]{@{}c@{}}Minimum\\ CR\end{tabular} &
  \begin{tabular}[c]{@{}c@{}}Maximum\\ CR\end{tabular} \\ \hline
Run-length & 8 &  4.0 & 0.002 & 3.9 & 4.0 \\ \hline
Run-length & 16 &  7.8 & 0.01 & 7.6 & 7.8 \\ \hline
Zeromask & 8 &  7.9 & 0.003 & 7.87 & 7.94 \\ \hline
Shuffled Zeromask & 16 &  8.9 & 0.004 & 8.8 & 8.9 \\ \hline
\end{tabular}
\caption{Compression ratios of various encoding schemes---XPCS dilute.}
\label{table:crxpcsd} 
\end{center}
\end{table}

\begin{table}[]
\begin{center}
\begin{tabular}{|l|c|c|c|}
\hline
\multicolumn{1}{|c|}{\textbf{Input Size}} &

  \begin{tabular}[c]{@{}c@{}}Best\\ CR\end{tabular} &
  \begin{tabular}[c]{@{}c@{}}Worst\\ CR\end{tabular}  \\ \hline
8  &  0.35 &  0.29   \\ \hline
16 &  0.70 &  0.44  \\ \hline
\end{tabular}
\caption{Small input sizes with gzip.}
\label{table:smallgzip} 
\end{center}
\end{table}


\subsubsection{Design methodology using Chisel}

\begin{figure}[htbp]
\centering 
\includegraphics[width=1\textwidth,origin=c,angle=0]{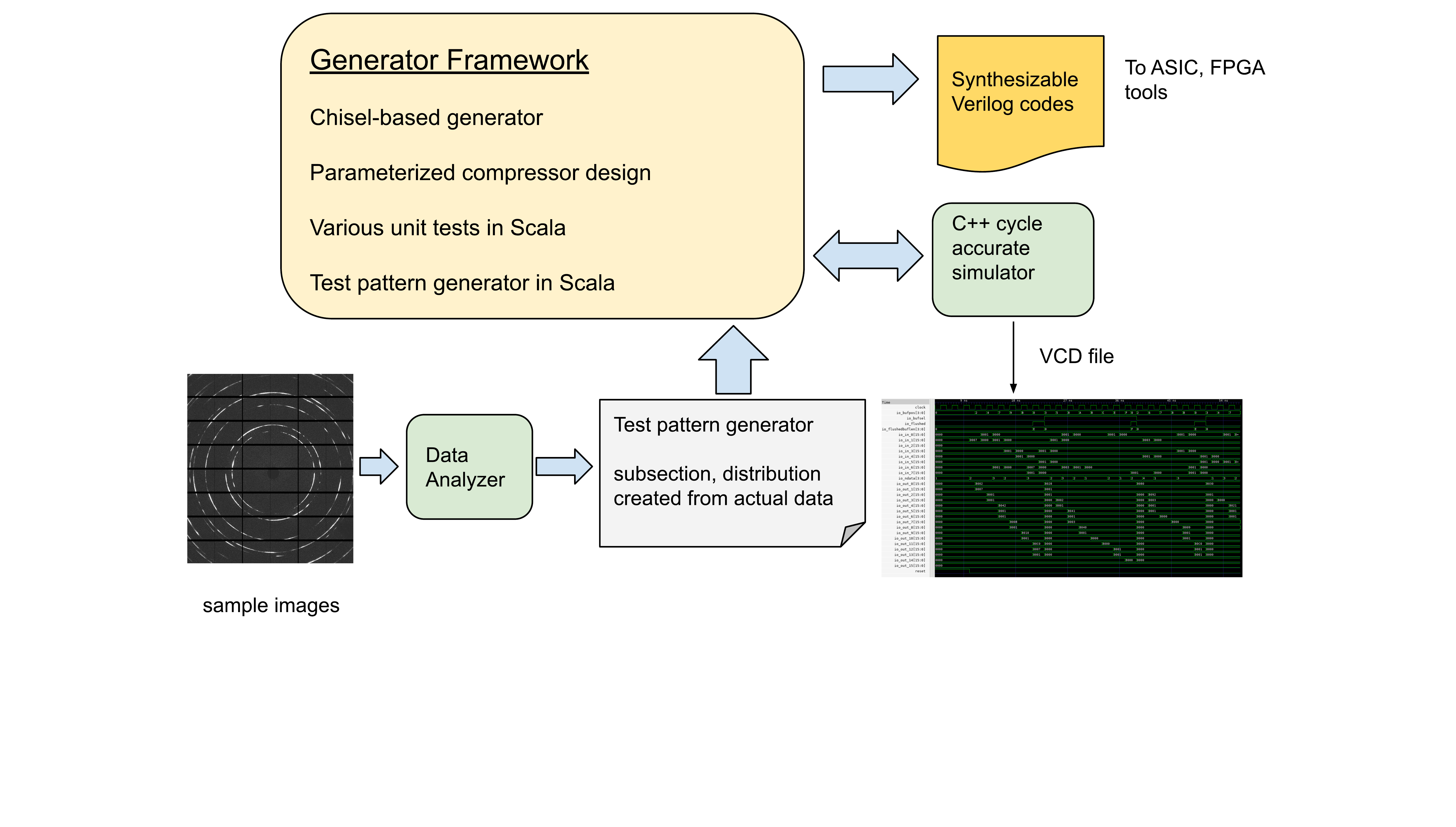}
\caption{Hardware compressor generator framework.}
\label{hacogen} 
\end{figure}

Before we describe the design and implementation of the streaming edge compressor, we describe our design methodology.
To improve both the productivity and quality of digital circuit designs, 
we have designed and implemented a hardware compressor generator framework (Figure \ref{hacogen}) that generates a concrete compressor RTL design.

Our compressor design is highly parameterized and fully written in the emerging hardware design language Chisel \cite{Bachrach:ft} so that it can generate a wide range of hardware configurations (e.g., different numbers of input/output pixels and bit widths and eventually different compressor algorithms). The framework not only provides a mechanism to generate synthesizable Verilog codes of a streaming compressor with specified design parameters such as input/output sizes and  bit width; it also provides a fully integrated, flexible test harnesses and a test pattern generator that allows us to test the compressor design with various test patterns before Verilog is synthesized and implemented in a particular ASIC technology node. 

Chisel is rapidly gaining popularity in digital circuit design and has been used in various ASIC and FPGA projects. A notable project based on Chisel is the Rocket Chip generator \cite{Asanovic:EECS-2016-17}, which has generated RISC-V processor RTL designs with several tape-outs. Chisel is an open-source hardware construction language, implemented as class libraries of the Scala language. With the power of Scala, a modern functional, object-oriented language, it allows designers to express complicated circuit blocks far more easily and more concisely than hardware description languages (HDLs) allow; this improves readability and reduces errors. The important characteristics of Chisel are zero-cost abstraction, higher level of expressibility, and controllability. Chisel is not a high-level synthesis tool. Chisel not only provides the abstraction of digital circuit primitives such as combinational logics, wires, registers, and memories  but also provides some basic structures such as multiplexer, first-in first-out, queue, and shift registers, which are defined in the Chisel standard library. Additionally, larger building blocks such as floating-point operations are available as open source. Chisel also includes fully integrated frameworks for unit tests so that developers can write unit tests in Scala, and it seamlessly integrates external Verilog simulators such as Verilator. More important, it is fairly easy to install and use; a simple command line invocation of Chisel will generate Verilog codes, run Scala testbenches, and invoke Verilator to generate a value change dump (VCD) file.

\subsubsection{Design of streaming edge compressor}
\label{sec:compressor-design}
The streaming edge compressor is a data compression logic between the pixel array and a peripheral logic such as a network interface or a memory. The two most important design considerations of the streaming compressor are being stall-free (i.e., deterministic with a fixed latency) and having a small resource footprint. Since the pixel array generates (column) pixel data at a fixed rate, the compression logic must process the data without any stall condition, which would cause data dropping. For this design we avoided inferring a large-enough memory for the compression logic to temporarily store pixel data because of two factors: (1) the length of each experiment, which determines the size of a temporary memory, is unknown, and (2) the resources (e.g., the number of transistors) are scarce on the detector ASIC chip.

\begin{figure}[htbp]
\centering
\includegraphics[width=1\textwidth,origin=c,angle=0]{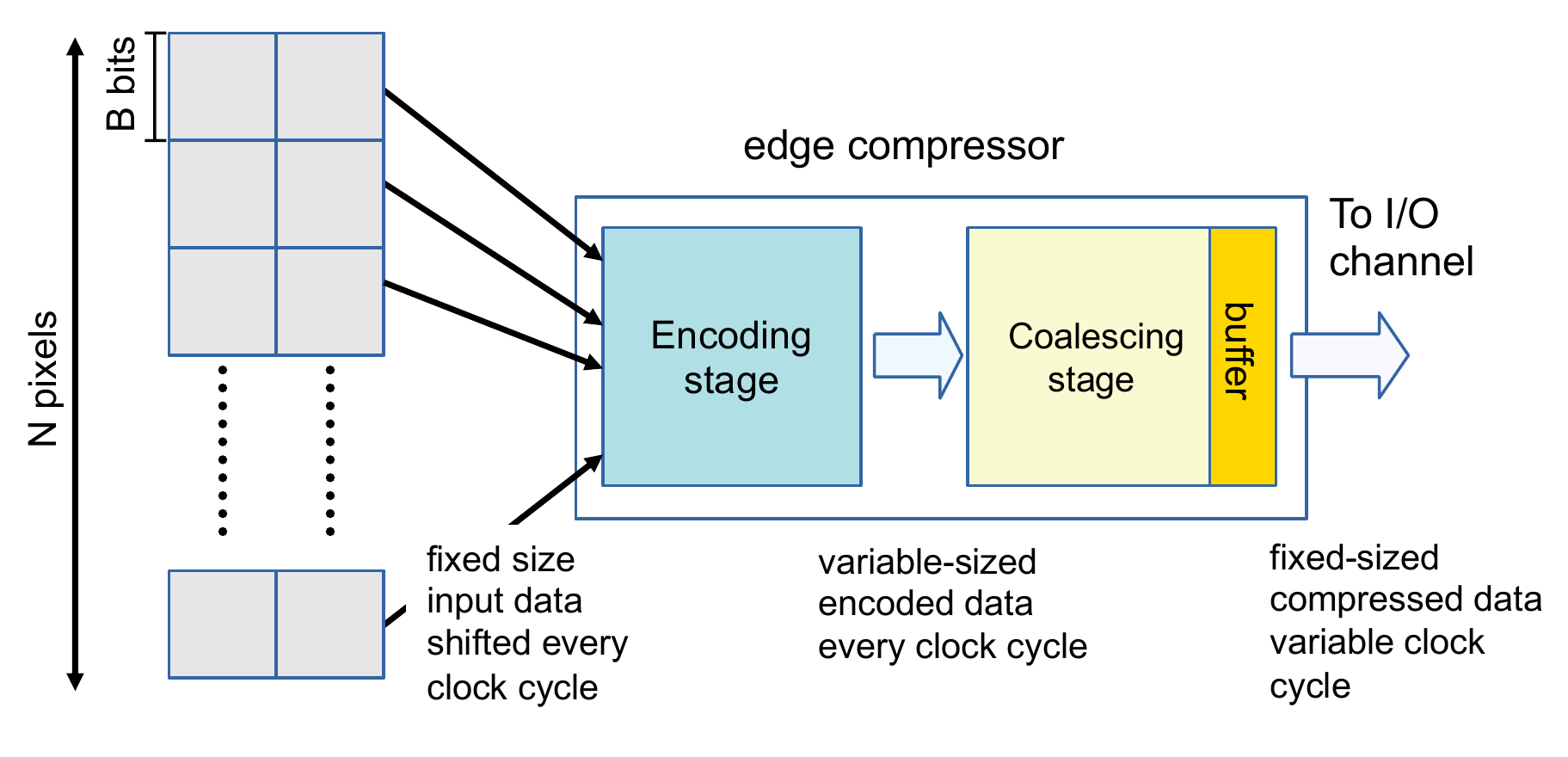}
\caption{Basic architecture of the streaming edge compressor}
\label{fig:hwcomp} 
\end{figure}

Figure~\ref{fig:hwcomp} shows the basic architecture of the streaming edge compressor. The main part of the streaming compressor is the encoding stage, which is an implementation of the zeromask compressor. 
The coalescing stage is needed for fixed-size peripherals (such as SRAMs, memory controllers). The encoding stage generates encoded data every clock cycle. The size of the encoded data varies depending on the content of the input data. The coalescing stage packs these variable-sized encoded data fragments into an internal buffer whose size aligns with the data bus width of a target device before sending it to the target device.  The number of clock cycles required until the internal buffer becomes ready to be read is also variable.

\paragraph{Encoding stage} The encoding stage receives column pixel data from the pixel array every clock cycle, performs the ZM encoding, and generates encoded data whose size is variable between 1 to (N+1), where N is the input size. 

\begin{figure}[htbp]
\centering 
\includegraphics[width=1\textwidth,origin=c,angle=0]{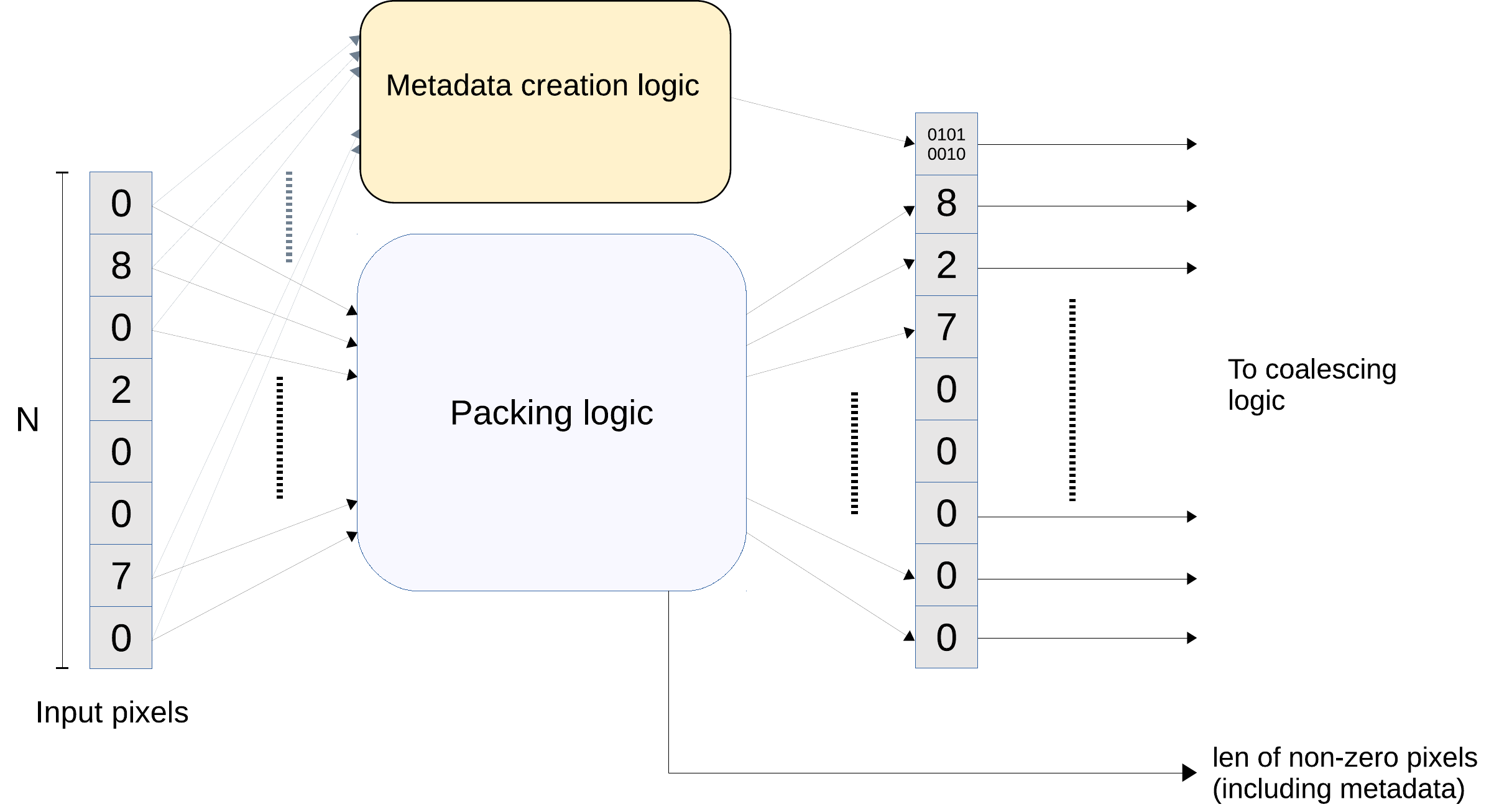}
\caption{Encoding stage: metadata creation and packing.}
\label{encodingstage} 
\end{figure}

Figure \ref{encodingstage} shows the basic idea of how the encoding stage generates an encoded data fragment. An encoded data fragment consists of a metadata header and a series of nonzero pixels; and the number of pixels encoded is chosen so that the metadata, a bit
field where a 0 bit encodes a suppressed pixel, fits a single pixel bit-length.
The metadata creation logic is basically a combination of comparators and bit-shift operations. Chisel allows such circuits to be expressed in a concise manner.

%

The rest of the encoding logic is the packing logic that removes zero pixels from the input pixel data, which includes both nonzero and zero pixels, and generates a sequence of pixel data only with nonzero pixels while preserving the order of the pixels. 



Internally the packing logic consists of N copies of a logic named \lstinline|ShiftUp|. Each \lstinline|ShiftUp| logic receives pixel data (N pixels) and the \lstinline|pos| indicator that holds the index of the current pixel position. Then it outputs updated pixel data and \lstinline|pos| that holds the index of the pixel for the next logic. All \lstinline|ShiftUp| logics are sequentially connected; the outputs of the previous logic are connected to the inputs of the next logic, except the first and last logic. The input to the first \lstinline|ShiftUp| logic receives pixel data from the pixel array. The initial value of \lstinline|pos| is set to zero. The outputs from the last \lstinline|ShiftUp| logic are connected to the coalescing stage, which includes the pixel data that hold packed pixels and the position index that is equal to the number of the nonzero pixels.


The number of lines in the entire packing logic is approximately 50, including boiler plate code, while the number of lines in the generated Verilog code is approximately 500 with \lstinline|npixel| = 8 and 1,600 with \lstinline|npixel| = 16. Note that the number of lines in the Chisel code stays the same for both configurations because it is fully parameterized. 

\paragraph{Coalescing stage} The data generated from the encoding stage are variable in size. If the interface of a peripheral device accepts variable-sized data, we can simply send encoded data into such a device. Otherwise, if the interface to the serializer expects a fixed size (e.g., 256 bits), a buffering mechanism is needed. The coalescing logic is a simple buffering logic that packs multiple variable-size encoded data into a fixed-size buffer before sending to the peripheral device.

\begin{figure}[htbp]
\centering 
\includegraphics[width=1\textwidth,origin=c,angle=0]{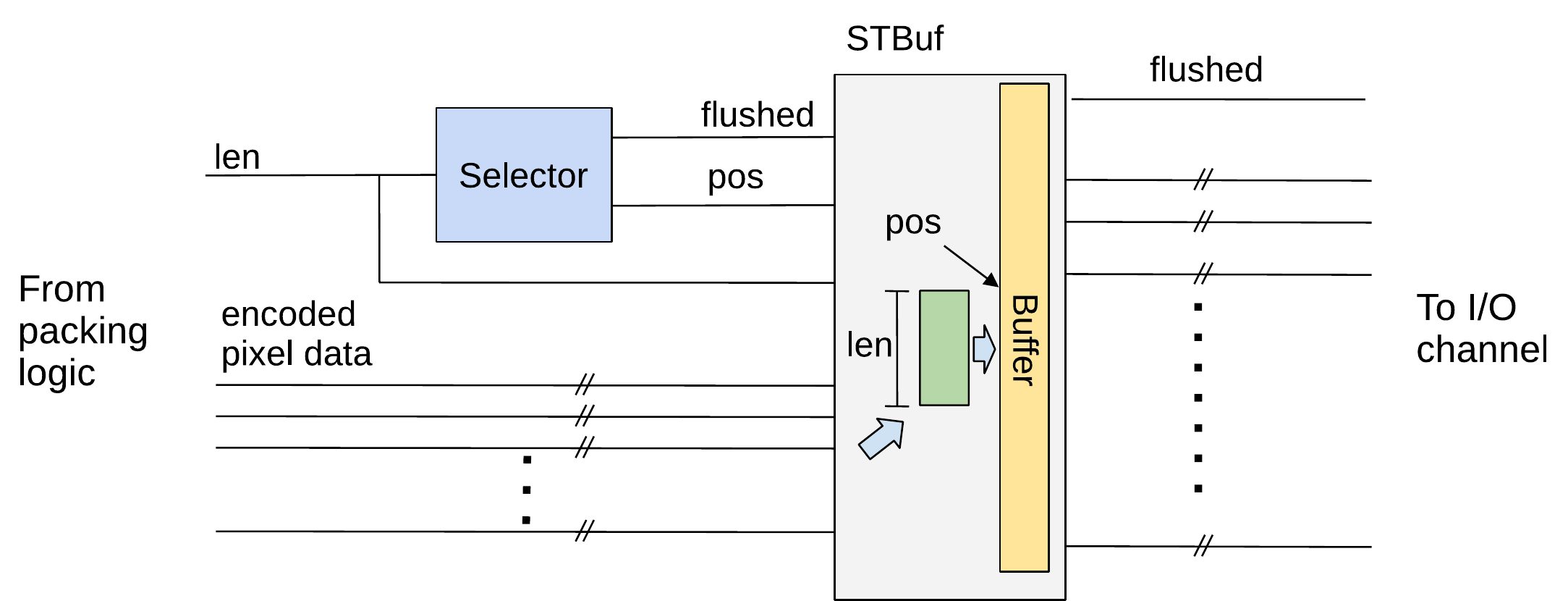}
\caption{Block diagram of the coalescing logic that packs encoded data into a fixed-size buffer.}
\label{coalescing} 
\end{figure}

The coalescing logic consists of two components: the \lstinline|Selector| and \lstinline|STBuf| module (see Figure \ref{coalescing}). The \lstinline|Selector| module receives the length of the encoded data every clock cycle from the encoding logic and updates the insert position (\lstinline|pos| in \lstinline|STBuf|) to the buffer in the \lstinline|STBuf| module of the associated encoded data. It raises the flushed signal when the buffer becomes full.  The \lstinline|STBuf| module receives \lstinline|flushed| and \lstinline|pos| from the \lstinline|Selector| module and the encoded data from the encoding stage and outputs the fixed-size data. It consists of an array of registers (buffer) and a barrel shifter.

\subsubsection{Verification}
\begin{figure}[htbp]
\centering
\includegraphics[width=1\textwidth,origin=c,angle=0]{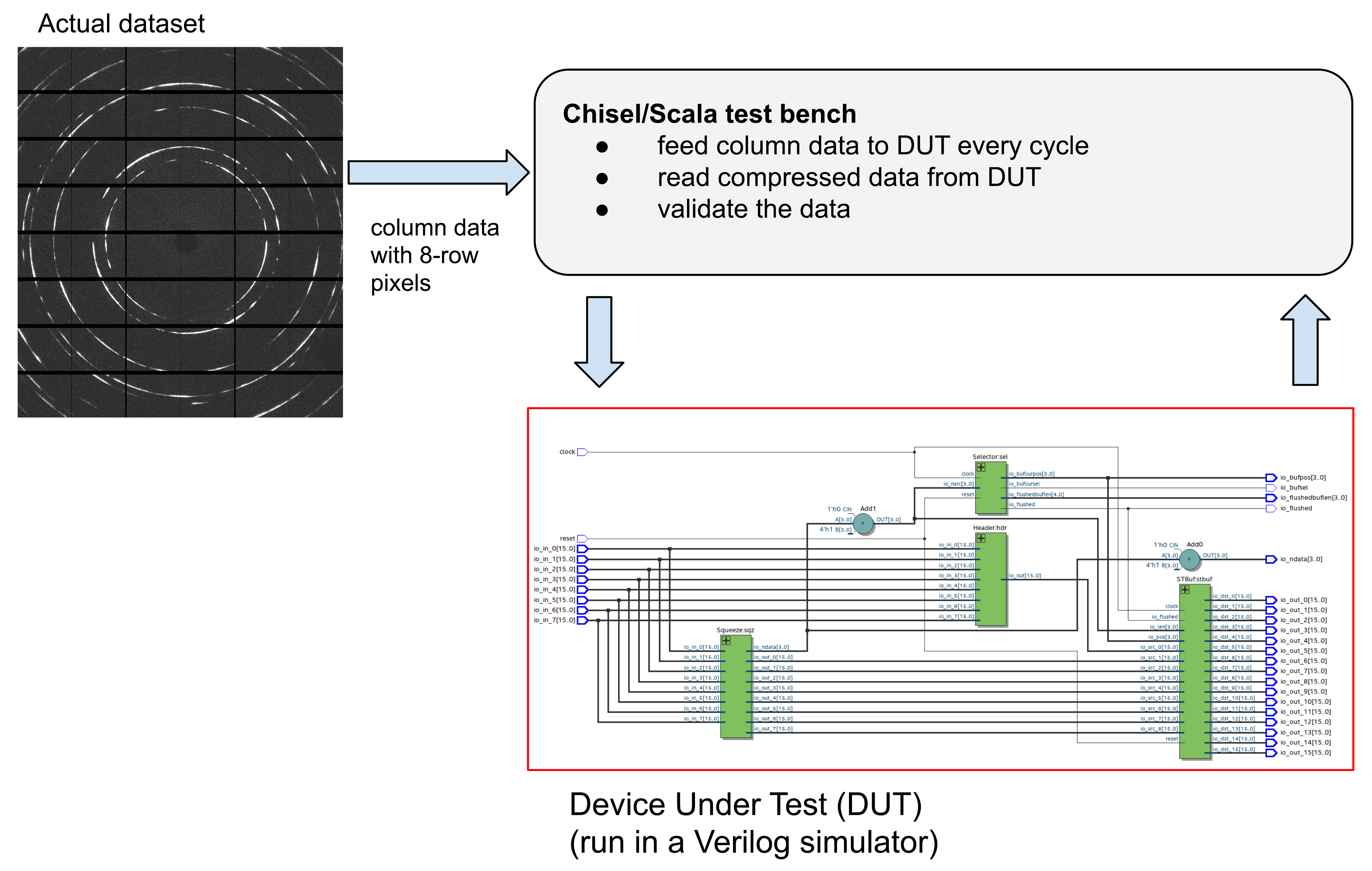}
\caption{Verification testbench.}
\label{testbench} 
\end{figure}

All of the design components for the streaming ZM compressor are expressed in Chisel and are translated into Verilog codes. While we could simply use an external Verilog simulator such as ModelSim to verify the functionality of the generated Verilog codes, Chisel provides a fully integrated testing harness for functional verification that allows developers to write testbenches in Scala, instead of Verilog. It seamlessly invokes an external Verilog simulator (Verilator by default). Since Scala is a powerful general-purpose functional, object-oriented programming language, it allows one to write complex and flexible testbench codes.
In addition we have successfully verified the functionality of the entire compressor design (Figure \ref{testbench}) against all four X-ray datasets. The testbench reads data from actual datasets and uses the data as input  to the device under test (DUT) that runs a simulated streaming compressor. An image frame in a dataset is chunked up into N-row pixel strips, which are fed into DUT column by column. We iterate all frames in a dataset. For example, with N=16, a 256x256 image frame is chunked up to 16 strips,\footnote{Each strip is each compressor instance.} and each strip has 256 columns. The total encoding count per frame is 4,096 ($= 16 \times 256$) in this example.

\subsubsection{Physical implementation}
\begin{figure}[htbp]
\centering 
\includegraphics[width=1\textwidth,origin=c,angle=0]{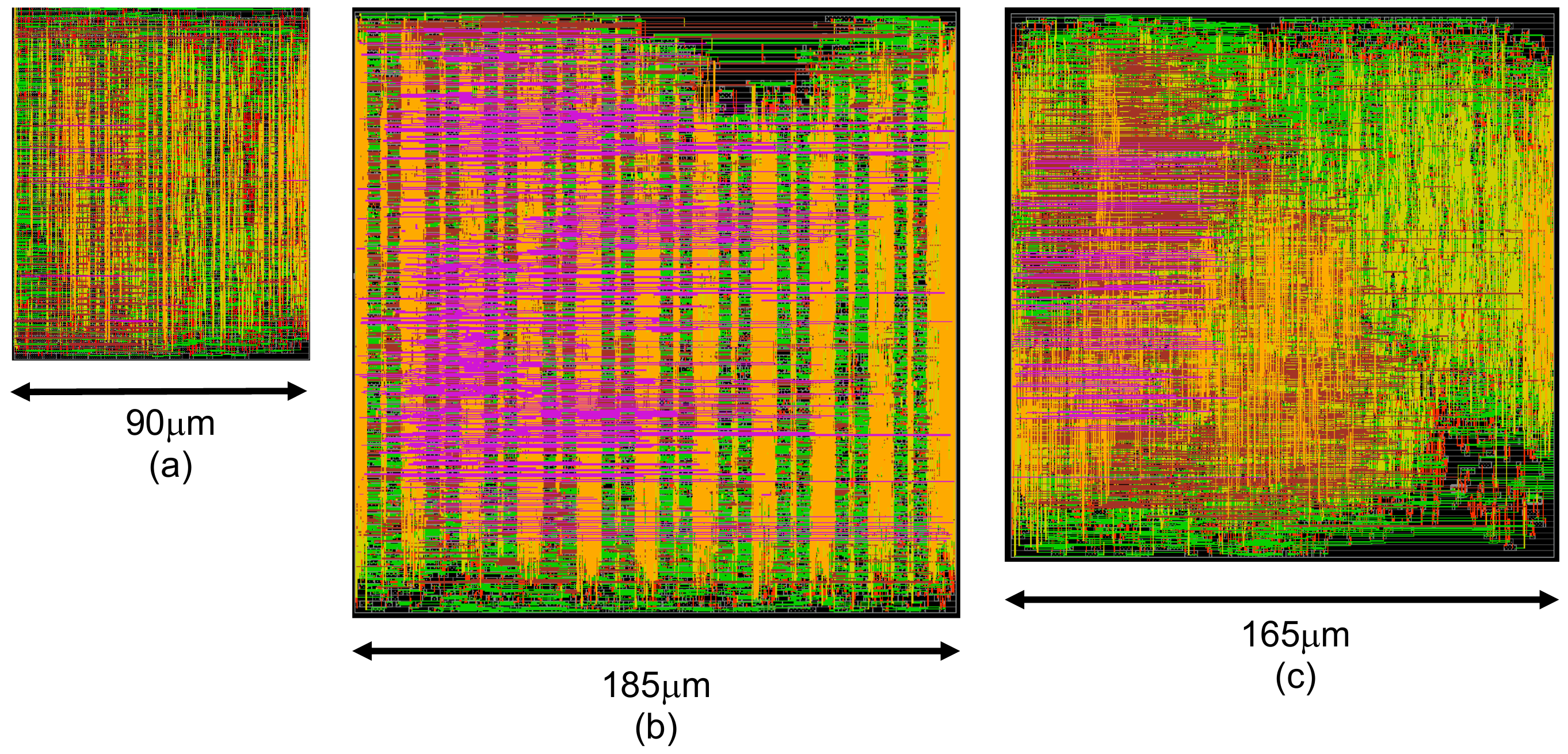}
\caption{Physical layout of three variants of the ZM compressor in \SI{65}{\nm} CMOS: (a) 8-pixel, (b) 16-pixel,  and (c) 16-pixel bit shuffled.}
\label{compressor-layout} 
\end{figure}

Figure \ref{compressor-layout} shows examples of the described streaming compressors that were coded in RTL, synthesized, and placed and routed in \SI{65}{\nm} standard cell CMOS. Layout (a) shows a placed and routed 8-pixel ZM compressor.
The area of this block is close to the desired area of a single pixel, which means that an 8-pixel streaming compressor block could be located at the end of every pixel row with little impact on the ASIC floor plan. Placing one compressor per pixel row would add the equivalent area of a single column of pixels to the pixel array. Layout (b) shows a placed and routed 16-pixel ZM compressor. The area of this block would span two \SI{100}{\um} pixel rows. If one 16-pixel compressor block were placed at the end of every two pixel rows, it would add the equivalent area of 2 pixel columns to the pixel array.
Since the compressor block of layout (a) supports 8 pixel inputs, 8 parallel shift buses can be implemented per row, resulting in an 8-fold reduction in the row shift time.
Layout (c) shows a 16-pixel bit-shuffled ZM compressor. This version is slightly smaller than the non-shuffled example of (b). 

\section{Outlook and Conclusions}
Two compression methods have been described that reduce the number of bits required to be streamed from a detector ASIC. The in-pixel encoding described in Section \ref{sec:in-pixel-encoding} demonstrates a compression ratio of  $>\!1.5\times$ independent of the dataset. On the other hand, the streaming edge compressor described in Section \ref{sec:edge-compression} depends on the nature of the X-ray dataset, in particular on the spatial distribution of zero-valued pixels for the shuffled ZM scheme. For example, the ptychography dataset has the worst compression ratio near the center of the detector, while the XRD dataset has the worst compression ratio near the strong Bragg peaks. These results have implications for the system-level design of detectors for multiple ASIC with on-chip compression. However, let us consider the impact on the off-chip bandwidth required for a hypothetical detector ASIC for the worst-case compression ratios (i.e., $3.5\times$, $6.9\times$, $8\times$, and $8.8\times$ for the XRD, ptychography, XPCS concentrated, and dilute, respectively). When combined with the in-pixel compression, the total worst-case compression ratios range from 5.25 to 13.2$\times$. If a $256\times256$ pixel array running at a \SI{1}{\MHz} frame rate is assumed and if each pixel has a 12-bit ADC with 2 gain bits, the uncompressed transmission bit-rate would be 917\,Gbps. Individual high-speed transmitters in \SI{65}{\nm} CMOS have been demonstrated at 5--10\,Gbps \cite{Chen_2019,Moreira} and the Timepix4 ASIC is designed with $16\times10.24 = 164\,$Gbps off-chip bandwidth \cite{Timepix4}. At these extreme bit-rates the number of transmitters needed to send uncompressed data would make a \SI{1}{\MHz} frame rate out of the question. With the worst-case compression ratios, the required off-chip bandwidth is reduced to 70--175\,Gbps. This can more reasonably be achieved by providing 7--18 10\,Gpbs transmitters at the edge of the detector ASIC, for example. Conversely, we can imagine detector systems with a variable frame rate for different X-ray datasets.  

In this paper we focused on a compression scheme that leverages statistical redundancy (higher zero occurrences) in single-cycle input pixels from the pixel array that allows us to design a simple yet effective compressor. As a next step we plan to explore compression algorithms and circuit designs that can exploit further statistical redundancy in spatial and temporal direction dimensions in order to improve the compression ratio further.

\acknowledgments
We thank Andrew Chihpin Chuang, Junjing Deng, and Suresh Narayanan for the X-ray datasets. We thank Gail Pieper for editing this manuscript. We thank Franck Cappello for discussions on compression schemes. We thank Pete Beckman and Alec Sandy for encouraging this multidisciplinary collaboration between the X-ray Science Division (XSD) and the Mathematics and Computer Science (MCS) Division. We also thank two anonymous referees for their  useful comments. This work is based in part on work supported by the U.S. Department  of Energy, Office of Science, under contract  DE-AC02-06CH11357. This research used resources of the Advanced Photon Source, a U.S. Department of Energy (DOE) Office of Science User Facility operated for the DOE Office of Science by Argonne National Laboratory under Contract No. DE-AC02-06CH11357. This research also used the resources of the Fermi National Accelerator Laboratory (Fermilab), a U.S. Department of Energy, Office of Science, HEP User Facility. Fermilab is managed by Fermi Research Alliance, LLC (FRA), acting under Contract No. DE-AC02-07CH11359.

\bibliographystyle{ieeetr}
\bibliography{references}
\end{document}